\documentclass[twocolumn,showpacs,preprintnumbers,nofootinbib]{revtex4}

\usepackage{graphicx}
\usepackage{dcolumn}
\usepackage{bm}
\usepackage{amssymb}
\usepackage{amsmath}
\usepackage{epsfig}
\usepackage[dvips]{color}    

\def\beq{\begin{equation}}
\def\eeq{\end{equation}}
\def\bea{\begin{array}}
\def\eea{\end{array}}
\def\be{\begin{equation}}
\def\ee{\end{equation}}
\def\ba{\begin{eqnarray}}
\def\ea{\end{eqnarray}}

\def\[{\left[}
\def\]{\right]}
\def\({\left(}
\def\){\right)}

\def\sm0{{\widetilde{m}_0}}

\def\U1em{{U(1)_{\rm em}}}

\def\sq2{\sqrt{2}}

\def\ee{e^+e^-}

\def\End{\end{document}}

\def\fsl#1{\setbox0=\hbox{$#1$}                 
   \dimen0=\wd0                                 
   \setbox1=\hbox{/} \dimen1=\wd1               
   \ifdim\dimen0>\dimen1                        
      \rlap{\hbox to \dimen0{\hfil/\hfil}}      
      #1                                        
   \else                                        
      \rlap{\hbox to \dimen1{\hfil$#1$\hfil}}   
      /                                         
   \fi}  

\begin{document}                                                              

\title{%
Neutrino Masses from Loop-induced $d \geq 7$ Operators
}
  
\author{%
{\sc Shinya Kanemura\,$^1$ and Toshihiko Ota\,$^2$}
}

\affiliation{%
\vspace*{2mm} 
$^1$Department of Physics, University of Toyama, 
3190 Gofuku, Toyama 930-8555, Japan
\\
$^2$Max-Planck-Institut f\"{u}r Physik
(Werner-Heisenberg-Institut)\\
F\"{o}hringer Ring 6,
80805 M\"{u}nchen, Germany
\\
}

\preprint{%
MPP-2010-127,
UT-HET-047
}

\pacs{%
11.30.Fs,
12.60.Fr,
14.60.Pq,
14.60.St,
}

\begin{abstract}
We propose a new scenario where neutrino masses are generated 
via operators with the mass dimension higher than five, 
which are induced at the loop level.
The scenario is demonstrated with concrete models where neutrino masses 
are generated via a one-loop dimension-seven operator which 
is induced through TeV scale dynamics under the exact $Z_{2}$ symmetry.
Tiny neutrino masses are naturally induced 
from the TeV scale dynamics without introducing any artificial 
assumption on magnitudes of coupling constants.   
The combination of one-loop factor $1/(4\pi)^2$ and 
the factor of the ratio $(v/\Lambda)^2$ 
between the electroweak scale $v$
and new physics scale $\Lambda$
provides sufficient suppression as compared to the model based on 
the dimension-five operator induced at the tree level. 
The reproduction of the data for neutrino masses and mixings are
 discussed under the constraint from experiments 
for lepton flavour violation.
We also mention phenomenological implications at collider experiments 
and dark matter candidates.
\end{abstract}

\maketitle

\setcounter{footnote}{0}
\renewcommand{\thefootnote}{\arabic{footnote}}

\section{Introduction}

Mystery is the origin of tiny neutrino masses that are 
indicated from the neutrino oscillation data. 
How can we understand the smallness of neutrino masses 
as compared to the electroweak scale?
A simple way of the explanation may be based on the seesaw
mechanism~\cite{Minkowski:1977sc,Yanagida:1979,GellMann:1979,Mohapatra:1979ia}, 
introducing 
right-handed neutrinos with large Majorana masses
at the scale such as that of grand unification.
Although this is an attractive scenario, 
introduction of such large masses causes 
another hierarchy among mass scales.
In addition, such a large mass scale is beyond the experimental
reach and the theory would be untestable directly.

In the Standard Model (SM),
the Higgs sector, on the other hand, is the last uncharted 
part.
Although the SM Higgs sector is the simplest scenario 
with a scalar isospin doublet, 
the true Higgs sector may take a non-minimal form. 
Such an extended Higgs sector 
may be closely related to the mechanism to induce tiny neutrino masses
at the TeV scale.
Such a possibility is interesting because the model 
is in principle testable directly at on-going and 
future collider experiments, 
such as the Fermilab Tevatron, the CERN Large Hadron Collider (LHC) 
and the International Linear Collider (ILC).

If neutrino masses are of the Majorana type, they are generated 
through the lepton number violating effective operators.
In the usual seesaw scenarios, the neutrino masses are derived 
from the dimension-five operator $\nu\nu\phi\phi/\Lambda$, 
where $\nu$ represents left-handed neutrinos, 
$\phi$ does the Higgs boson, and 
$\Lambda$ is a scale of the new physics.
In a class of models where neutrino masses are radiatively 
generated, such a dimension-five operator is induced at 
the loop level by the TeV scale dynamics. 
For example, in the model proposed by A.~Zee~\cite{Zee:1980ai,Zee:1985rj}, 
the dimension-five operator is generated at the one-loop level 
via the lepton number violating interaction and dynamics of 
the extended Higgs sector.
In the model proposed by E.~Ma~\cite{Ma:2006km}, 
the dimension-five operator is also 
generated at the one-loop level via the physics of the extra scalar 
doublet and the TeV scale right-handed neutrinos, where the both of 
new fields are assigned odd quantum number under the discrete 
$Z_2$ symmetry.
Such a one-loop generation of neutrino masses from the TeV 
scale dynamics, however, still requires unnaturally small 
coupling constants for reproducing the tiny neutrino masses.
There are several models in which neutrino masses are generated at 
the two-loop level~\cite{Zee:1985id,Babu:1988ki,Aoki:2010ib,Babu:2010vp} 
and also the three-loop level~\cite{Krauss:2002px,Cheung:2004xm,Aoki:2008av,Aoki:2009vf},
where such fine tuning is not necessary because of the sufficient 
suppression by additional loop factors.
In all these models, dimension-five operators are 
induced at the loop level.

Recently, a new idea has been proposed where 
tiny neutrino masses are generated via the operators whose 
dimension is higher than 
five~\cite{Gogoladze:2008wz,Babu:2009aq,Bonnet:2009ej,Picek:2009is,Liao:2010rx}.
In Ref.~\cite{Bonnet:2009ej}, 
some concrete examples are examined, 
in which neutrino masses are generated 
via the dimension-seven operator $\nu\nu\phi\phi\phi\phi/\Lambda^3$ 
which are induced at the tree level with the extend scalar dynamics. 
In this case, there is an additional suppression factor of 
$(v/\Lambda)^2$ as compared to neutrino masses 
via the dimension-five operators, where $v$ ($\simeq 246$ GeV) 
is the vacuum expectation value (vev) of the Higgs boson.
Although these models are interesting, 
a sort of fine tuning is still required especially 
to reproduce the scale of neutrino masses, 
when $\Lambda$ is assumed to be of TeV scale.
%

\begin{figure*}[htb]
\unitlength=1cm
\begin{picture}(16,3)
\put(0,0){\includegraphics[width=7.5cm]{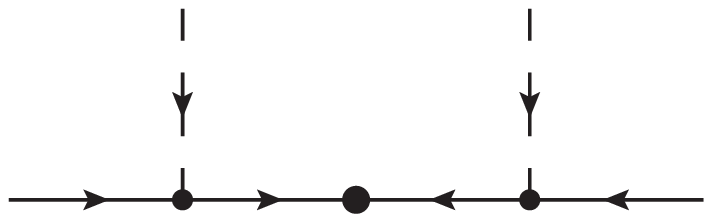}}
\put(8,0){\includegraphics[width=7.5cm]{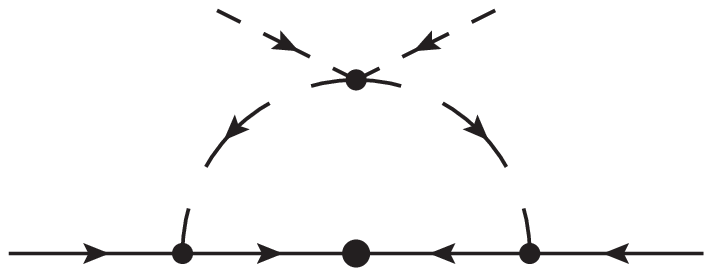}}
\put(0,0){$L$}
\put(7.2,0){$L$}
\put(1.9,2.5){$H_{2}$}
\put(5.4,2.5){$H_{2}$}
\put(3.5,-0.2){$N_{R}$}
\put(8,0){$L$}
\put(15.2,0){$L$}
\put(9.9,2.8){$H_{2}$}
\put(13.3,2.8){$H_{2}$}
\put(10,1.5){$\eta$}
\put(13.3,1.5){$\eta$}
\put(11.5,-0.2){$N_{R}$}
\thicklines
\put(7.2,1){\vector(1,0){1}}
\end{picture}
\caption{Schematic explanation of the method to make 
a loop digram from the tree diagram for neutrino masses.
The Higgs doublets $H_{2}$ in the tree seesaw diagram (left)
are substituted by the inert doublets $\eta$ with odd parity, 
and the loop of the inert doublet is closed 
by the quartic coupling of $(\eta^{\dagger} H_{2})^{2}$.
This loop dimension-five model (right) 
was proposed in Ref.~\cite{Ma:2006km}.
Here $N_{R}$ represents right-handed neutrinos.}
\label{Fig:seesaw-to-Ma}
\end{figure*}

In this paper, we propose a scenario in which neutrino masses are 
generated via higher-dimensional operators 
$\nu\nu (\phi \phi)^{(d-3)/2}/\Lambda^{d-4}$ 
($d = 7,9,11 \cdots$) 
which are induced by quantum effect.
In general, the size of neutrino masses 
from the operator with the mass dimension $d$, 
which arises from a $n$-loop diagram, 
can be estimated as
\begin{align}
m_{\nu}
\sim
v
\times
\left(
\frac{1}{16\pi^{2}}
\right)^{n}
\times
\left(\frac{v}{\Lambda}\right)^{d-4}.
\label{eq:mNu-symbolic}
\end{align}
In the models with $d=7$ and $n=1$, 
neutrino masses are further suppressed by 
the one-loop factor $1/(16 \pi^{2})$ 
and the factor $(v/\Lambda)^2$ 
as compared to the tree-induced dimension-five operator case
(=the ordinary seesaw model).
%
In such models,
the new physics scale $\Lambda$ may be set on the TeV scale
without assuming any unnatural small coupling constant.
%
In order to realize this scenario, we impose 
an exact $Z_2$ parity~\cite{Krauss:2002px,Ma:2006km} 
and an approximate discrete 
symmetry~\cite{Bonnet:2009ej,Giudice:2008uua} 
to forbid the appearance of the dimension-five operator 
as well as 
the dimension-seven operators induced at the tree level. 
In such models, the lightest $Z_{2}$ odd particle can be a  
Dark Matter (DM) candidate as long as it is electrically 
neutral.

We show two concrete examples of the models along this line.
It is demonstrated that the models can reproduce the neutrino 
data for the masses and mixings without fine tuning among 
coupling constants due to the TeV scale dynamics of the models.  
We discuss the constraint on parameters of the models from 
the data of lepton flavor violation~\cite{Amsler:2008zzb,Adam:2009ci}. 
In these models, extended scaler sectors appear 
with the exact $Z_2$ symmetry, 
which provide rich phenomenological predictions. 
We mention the test of the models at current and future collider 
experiments at the LHC and the ILC.

This paper is organized as follows:
In Sec.~\ref{Sec:method}, we briefly recapitulate the method to realize
the higher dimensional neutrino mass generation with 
an approximate discrete symmetry. 
We also review the way to make a tree diagram for neutrino masses 
become the loop diagram by introduction of $Z_{2}$ parity. 
Combining with these two methods, we construct two concrete models
in Sec.~\ref{Sec:models},
in which neutrino masses arise from the effective 
dimension-seven operator which is induced at the one-loop level.
In Sec.~\ref{Sec:Summary}, 
we discuss some phenomenological aspects of the models.

\section{Method}
\label{Sec:method}

Before we come on to descriptions of the concrete models,
let us look briefly at the essentials for 
the tree-level dimension-seven neutrino mass 
generation~\cite{Bonnet:2009ej}.
There are two key components to produce the effective 
dimension-seven operator for neutrino masses
at the electroweak scale:
\begin{itemize}
\item An additional symmetry to forbid the dimension-five 
      $\nu\nu \phi\phi/\Lambda$ operator.
      The simplest choice for non-supersymmetric models is $Z_{5}$.
 
\item The extended Higgs sector with two Higgs doublets
      so that the combination $(H_{1} H_{2})$ 
      can carry a charge under the additional symmetry.
      Here the hypercharge of $H_{1}$ is given to be $-1/2$ and  
      that of $H_{2}$ is $+1/2$.
\end{itemize}
Taking the setups and assigning appropriate charges
to the standard model particles, 
we can forbid the dimension-five operator and make
\begin{align}
\mathcal{L}_{\text{eff}}
=
\frac{\mathcal{C}}{\Lambda^{3}}
LLH_{2} H_{2} H_{2} H_{1}
\label{eq:effop-dim7}
\end{align} 
to be the leading contribution to neutrino masses,
where $\mathcal{C}$ is a mass dimensionless coefficient\footnote{%
The choice of dimension-seven operators which contribute to
neutrino masses is not unique~\cite{Bonnet:2009ej}. 
In this paper, we concentrate on the operator shown 
in Eq.~\eqref{eq:effop-dim7}. 
}.
The possible models for this tree-level dimension-seven
neutrino mass generation mechanism 
are listed in Ref.~\cite{Bonnet:2009ej}.
In the models including the SM singlet fermions
at the high energy scale,
one can see that the $Z_{5}$ symmetry forbids the fermions 
(=right-handed neutrinos) to have the Majorana mass term.
Because of the absence of the Majorana mass term, 
the dimension-five operator cannot arise at the electroweak 
scale and the dimension-seven operator dominates 
the contribution to neutrino masses.   

Extending these models, 
we consider the models,
in which neutrino masses are generated via the 
dimension-seven operator 
but the effective operator is induced through a one-loop 
diagram. 
To construct such loop-induced models, we follow the method with 
the exact $Z_{2}$ symmetry, which was developed 
in Refs.~\cite{Krauss:2002px,Ma:2006km}.
%
%
The essential is introduction of the inert doublet 
with the odd parity under the $Z_{2}$ symmetry.
Due to the exact $Z_{2}$ parity, the inert doublet 
cannot take a vacuum expectation value (vev).
Let us describe the maneuver, taking the ordinary type-I 
seesaw model as an example.
The procedure is schematically illustrated 
in Fig.~\ref{Fig:seesaw-to-Ma}.
Assigning the odd parity to the right-handed neutrinos $N_{R}$ 
and substituting the inert doublet $\eta$ for 
the Higgs doublet $H_{2}$ in the neutrino Yukawa interaction, 
one can forbid the 
tree-level contribution to neutrino masses.
The inert doublets in the diagram Fig.~\ref{Fig:seesaw-to-Ma}
are converted to the Higgs doublets 
through a quartic interaction,
\begin{align}
\mathcal{L} = \frac{\lambda}{2}
(\eta^{\dagger} H_{2}) (\eta^{\dagger} H_{2})
+{\rm H.c.}.
\end{align}
In other words, the inert Higgs legs are closed by the quartic 
interaction and make a loop.
This leads to the one-loop diagram for neutrino masses, 
which was proposed in Ref.~\cite{Ma:2006km}.
In the following sections, we will apply this procedure 
to the models in which neutrino masses are
generated through the dimension-seven operator which 
is induced via tree diagrams, 
and build the loop-induced dimension-seven models.

\section{Models}
\label{Sec:models}

We here consider two examples 
to illustrate the method to build the models in which 
neutrino masses are generated through the effective dimension-seven 
operator induced from a one-loop diagram.
 
\subsection{Model A}
The renormalizable models to induce the effective 
interaction Eq.~\eqref{eq:effop-dim7}
from tree diagrams at the electroweak scale are 
listed in Ref.~\cite{Bonnet:2009ej}.
In this subsection, we employ the model described 
as Decomposition~\#1 among them, in which the SM 
gauge singlet Dirac fermion $\psi$
and the singlet scalar $\varphi$ are introduced.
The particle contents and the charge assignments 
are summarized in Table~\ref{Tab:Decom1-charge}.
Here, the charges for the quarks and leptons 
are assigned so as to reproduce the Yukawa interactions 
of type-II two-Higgs-doublet model (THDM)\footnote{%
\label{footnote:THDM}
In general, there are four possibilities 
for the Yukawa interaction in THDM 
under the (softly-broken) discrete $Z_{2}$
symmetry~\cite{Barger:1989fj,Grossman:1994jb,Akeroyd:1994ga,Akeroyd:1996di,Akeroyd:1998ui,Aoki:2009ha,Su:2009fz,Logan:2009uf}. 
All the possibilities can also be realized with appropriate 
charge assignments in the case of the $Z_{5}$ symmetry.
}.
For detailed arguments for this model and the 
(softly broken) $Z_{5}$ symmetry, see Sec.~3.1 
in Ref.~\cite{Bonnet:2009ej}.   
\begin{table}[tb]
\begin{tabular}{ccccccccccc}
\hline \hline
& $L$ & $e^{c}$ & $Q$ & $u^{c}$ & $d^{c}$ & $H_{2}$ & $H_{1}$ 
& $\psi ({\bf 1}^{D}_{0})$ & $\eta ({\bf 2}^{s}_{1/2})$ 
& $\varphi({\bf 1}^{s}_{0})$
\\
\hline
softly broken $Z_{5}$
&
1 &1 &0 & 0 & 2 & 0 & 3 & 1 & 0 & 3 \\
exact $Z_{2}$
&
+ & + & + & + & + & + & +
&
$-$ & $-$ & +
\\
\hline \hline
\end{tabular}
\caption{Particle contents and charge assignments for 
Model A.
The symbol ${\bf X}_{Y}^{\mathcal{L}}$ indicates
the representations of the fields; 
${\bf X}$ for $SU(2)_{L}$, $Y$ for $U(1)_{Y}$,
and $\mathcal{L}$ for Lorenz group; i.e.,
Dirac spinor ($D$) and scalar ($s$).
}
\label{Tab:Decom1-charge}
\end{table}
In this letter, we are interested in the neutrino masses 
induced from the dimension-seven operator Eq.~\eqref{eq:effop-dim7}
but the effective interaction is realized by a loop diagram. 
In order to forbid arising the dimension-seven operator from 
a tree diagram, we introduce the exact $Z_{2}$ parity 
and an inert doublet $\eta$, and assign the odd charge 
for the inert doublet and the 
singlet Dirac fermion $\psi$.

\begin{figure}[tb]
\unitlength=1cm
\begin{picture}(8,5)
\put(0,0){\includegraphics[width=8cm]{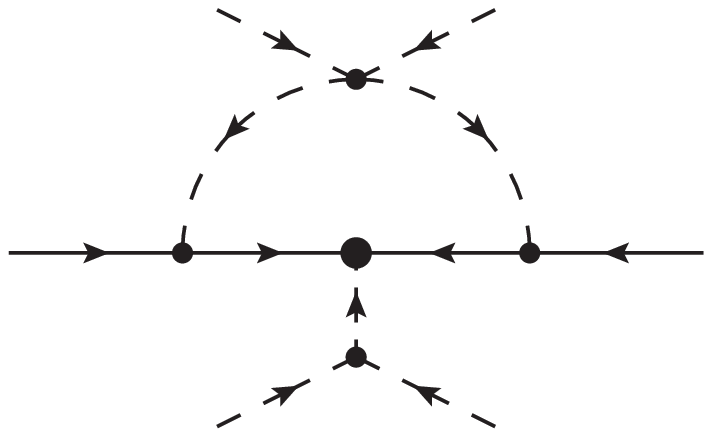}}
\put(0,2){$L$}
\put(7.8,2){$L$}
\put(2.6,1.5){$\psi$}
\put(5,1.5){$\psi$}
\put(4.2,1.3){$\varphi$}
\put(2,3){$\eta$}
\put(5.8,3){$\eta$}
\put(2.05,4.6){$H_{2}$}
\put(5.6,4.6){$H_{2}$}
\put(2.05,0){$H_{2}$}
\put(5.6,0){$H_{1}$}
\end{picture}
\caption{Diagram for neutrino masses in Model A.
The lepton number is violated at 
the interaction of $\psi$-$\psi$-$\varphi$,
which is shown with a fat blob.}
\label{Fig:mNu-Decom1}
\end{figure}

The Lagrangian of the fundamental interactions for the neutrino mass 
generation is given as
\begin{align}
\mathcal{L}
=&
\mathcal{L}_{\rm SM}
\nonumber
\\
&+
\biggl[
{(Y_{\nu})_{a}}^{\alpha} 
\overline{\psi}^{a} 
{\rm P}_{L} 
(\eta {\rm i} \tau^{2} L_{\alpha})
+
(\kappa_{L})^{ab} 
\varphi
\overline{\psi^{c}}_{a} 
{\rm P}_{L} 
\psi_{b}
\nonumber 
\\
&\hspace{0.5cm}+
(\kappa_{R})^{ab} 
\varphi
\overline{\psi^{c}}_{a} 
{\rm P}_{R} 
\psi_{b}
+
\mu \varphi^{*} 
(H_{1} {\rm i} \tau^{2} H_{2})
+
{\rm H.c.}
\biggr]
\nonumber 
\\
&
+
{M_{a}}^{b}
\overline{\psi}^{a} 
\psi_{b}
+
m_{\varphi}^{2}
\varphi^{*} \varphi
+
m_{\eta}^{2} (\eta^{\dagger} \eta)
\nonumber 
\\
&+
\left[
\frac{\lambda}{2}
(\eta^{\dagger} H_{2}) 
(\eta^{\dagger} H_{2})
+
{\rm H.c.}
\right]
-
\mathcal{V}_{\text{scalar}},
\label{eq:L-Decom1}
\end{align}
where $a$, $b$ and $\alpha$ represent
the flavour indices.
Let us first focus on neutrino masses which are 
our main concern, and we will take up some phenomenological
consequences of this model and the part 
$\mathcal{V}_{\text{scalar}}$ of the scalar potential later.
With the interactions shown in Eq.~\eqref{eq:L-Decom1},
the dimension-seven operator for neutrino masses 
is induced by the one-loop diagram described 
in Fig.~\ref{Fig:mNu-Decom1}, which is evaluated as
\begin{align}
\mathcal{L}_{\text{eff}}
=&
\frac{1}{(4\pi)^{2}}
 \frac{\lambda \mu}{m_{\varphi}^{2} m_{\eta}^{2}}
 {(Y_{\nu}^{\sf T})^{\alpha}}_{a}
{(Y_{\nu})_{b}}^{\beta}
 \nonumber 
 \\
&\times 
 \Bigl[
 (\kappa_{L})^{a b}
 \frac{M_{a} M_{b}}{m_{\eta}^{2}}
 \mathcal{I}(x_{a},x_{b})
 +
 (\kappa_{R})^{a b}
 \mathcal{J}(x_{a},x_{b})
 \Bigr]
 \nonumber 
 \\
 &\times
 (
\overline{L^{c}}_{\alpha}
{\rm i} \tau^{2} H_{2}
) 
(H_{2}
{\rm i} \tau^{2} L_{\beta})
(H_{1} {\rm i} \tau^{2} H_{2}),
\end{align}
where the functions $\mathcal{I}$ 
and $\mathcal{J}$ are defined as
\begin{align}
\mathcal{I} (x_{a},x_{b})
=&
 \frac{1}{(1-x_{a})(1-x_{b})}
\nonumber 
\\
&\hspace{-1.5cm}\times
\Biggl[
1
 +
 \frac{
 (1-x_{b}) x_{a} \ln x_{a}
 }{
 (x_{a} - x_{b})(1-x_{a})}
 -
 \frac{(1-x_{a}) x_{b} \ln x_{b}}{
 (x_{a} - x_{b})
 (1-x_{b})} 
\Biggr],
\\
\mathcal{J}
(x_{a},x_{b})
=&
 \frac{1}{(1-x_{a})(1-x_{b})}
\nonumber 
\\
&\hspace{-1.5cm}\times
\Biggl[
1
 +
 \frac{
 (1-x_{b}) x_{a}^{2} \ln x_{a}
 }{
 (x_{a} - x_{b})(1-x_{a})}
 -
 \frac{(1-x_{a}) x_{b}^{2} \ln x_{b}}{
 (x_{a} - x_{b})
 (1-x_{b})} 
\Biggr],
\end{align}
with $x_{a} \equiv M_{a}^{2}/m_{\eta}^{2}$.
We obtain neutrino masses
\begin{align}
(m_{\nu})^{\alpha \beta}
=
-\frac{v^{2}}{2} \sin^{2} \beta 
{(Y^{\sf T})^{\alpha}}_{a}
(M_{\text{eff}}^{-1})^{ab}
{(Y_{\nu})_{b}}^{\beta},
\label{eq:mNu-A}
\end{align} 
which are the same form as those derived from 
the ordinary type-I seesaw scenario.
{\it The effective mass} $M_{\text{eff}}$ 
{\it for the right-handed neutrinos} is given to be
\begin{align}
(M_{\text{eff}}^{-1})^{ab}
=&
\frac{1}{(4\pi)^{2}}
\frac{v^{2}}{2}
\sin 2\beta
\frac{\lambda \mu}{m_{\varphi}^{2} m_{\eta}^{2}}
\nonumber 
\\
&\hspace{-1.2cm}\times
\left[
(\kappa_{L})^{a b}
\frac{M_{a} M_{b}}{m_{\eta}^{2}}
\mathcal{I}(x_{a},x_{b})
+
(\kappa_{R})^{a b}
\mathcal{J}(x_{a},x_{b})
\right].
\label{eq:Meff-A}
\end{align}
Therefore, it is guaranteed that 
this model can reproduce all the features of the 
neutrino flavour in the canonical type-I seesaw model.
%
From the expressions Eqs.~\eqref{eq:mNu-A}
and~\eqref{eq:Meff-A},
it turns out that
neutrino masses of the order of one eV
is compatible with TeV scale masses for new fields 
without assuming extremely tiny couplings in the model.
One can expect that the collider experiments 
are accessible to those fields.

Let us discuss the scalar potential
and the softly-broken discrete symmetry in this model.
With the exact $Z_{5}$ symmetry,
the part of the scalar potential, which is only 
including $H_{1}$ and $H_{2}$ is described as  
\begin{align}
\mathcal{V}_{\text{THDM}}
=&
m_{1}^{2} |H_{1}|^{2}
+
m_{2}^{2} |H_{2}|^{2}
+
\frac{\lambda_{1}}{2}
|H_{1}|^{4}
+
\frac{\lambda_{2}}{2}
|H_{2}|^{4}
\nonumber 
\\
&+
\lambda_{3}
|H_{1}|^{2} |H_{2}|^{2}
+
\lambda_{4}
|H_{1} {\rm i} \tau^{2} H_{2}|^{2},
\end{align}
and it actually respects the global $U(1)$ symmetry
including $Z_{5}$~\cite{Bonnet:2009ej}.
If the $U(1)$ symmetry is spontaneously broken 
with the vevs of the Higgs doublets, 
it leads a Nambu-Goldstone boson. 
To dodge this problem, here we assume $Z_{5}$ is an approximate 
symmetry at the new physics scale $\Lambda$ and 
introduce an explicit and soft-breaking term of $Z_{5}$
\begin{align}
\mathcal{V}_{\text{soft}}
=
m_{3}^{2} H_{1} {\rm i} \tau^{2} H_{2} + {\rm H.c.},
\label{eq:Z5soft}
\end{align}
by setting the scale $m_{3}$ at the electroweak scale.
This term does not invoke the dimension-five operator at the tree level,
but at the loop level. The dimension-five contribution 
is only constructed through the dimension-seven operator
of Fig.~\ref{Fig:mNu-Decom1} by connecting outer 
legs of $H_{2}$ and $H_{1}$.
Therefore, setting $m_{3}$ is smaller than $\Lambda$
(but large enough to avoid the bound to the pseudo 
Nambu-Goldstone boson), we can keep the contribution 
sub-dominant against that arises from the original 
dimension-seven diagram.
Notice that by appropriate assignment of $Z_5$ charges
for right-handed quarks and charged leptons we can
have a Yukawa interaction without flavor changing neutral
current at the tree level (See also footnote \ref{footnote:THDM}).
Phenomenological constraints and implications to  
the scalar sector are mentioned in Sec.~\ref{Sec:Summary}
together with the other example which will 
be illustrated in the next subsection.

Before we turn to another example, 
let us briefly discuss constraints from lepton flavour 
violation in this model.
Since the Higgs fields do not mediate 
flavour changing neutral currents at the tree level, 
the leading contribution to the lepton flavour 
violating processes arises from a diagram
with a loop between two Yukawa interactions.
The contribution is exactly the same as that in the original 
dark doublet model~\cite{Ma:2006km}, which was calculated 
in Ref.~\cite{Ma:2001mr,Kubo:2006yx}:
\begin{align}
{\rm Br}(\mu \rightarrow e \gamma)
=
\frac{3 \alpha_{\rm em}}{64 \pi (G_{F} m_{\eta}^{2})^{2}}
\left|({\mathcal{C}_{\text{A}})_{e}}^{\mu}\right|^{2},
\label{eq:Brmueg-Decom1}
\end{align}
where the mass-dimensionless coefficient 
$\mathcal{C}_{\text{A}}$
for Model A is given as 
\begin{align}
{(\mathcal{C}_{\text{A}})_{e}}^{\mu}
=&
{(Y_{\nu}^{\dagger})_{e}}^{a}
\mathcal{F}(x_{a})
{(Y_{\nu})_{a}}^{\mu},
\end{align}
and the function $\mathcal{F}$ is 
\begin{align}
\mathcal{F}(x_{a})
\equiv
\frac{1- 6 x_{a} + 3 x_{a}^{2} + 2 x_{a}^{3} - 6 x_{a}^{2} \ln x_{a}}
{6 (1-x_{a})^{4}}.
\end{align}
We can see that 
it might be essential to assume a large 
value for $m_{\eta}$
enough to avoid a sizable
$\mu \rightarrow e \gamma$ effect.
An alternative way to circumvent the large LFV process
is discussed in Ref.~\cite{Kubo:2006yx}.


%
%
%
%
%

\subsection{Model B} 

\begin{table}[tb]
\begin{tabular}{ccccccccccc}
\hline \hline
& $L$ & $e^{c}$ & $Q$ & $u^{c}$ & $d^{c}$ & $H_{2}$ & $H_{1}$ 
& $\psi ({\bf 1}^{D}_{0})$ & $\eta ({\bf 2}^{s}_{1/2})$ 
& $\eta'({\bf 2}^{s}_{-1/2})$
\\
\hline
soft br. $Z_{5}$
&
1 &1 &0 & 0 & 2 & 0 & 3 & 1 & 0 & 2 \\
exact $Z_{2}$
&
+ & + & + & + & + & + & +
&
$-$ & $-$ & $-$
\\
\hline \hline
\end{tabular}
\caption{Particle contents and charge assignments 
for the softly broken $Z_{5}$ and the exact $Z_{2}$ 
in Model B.}
\label{Tab:Decom13-charge}
\end{table}

Let us show the second example with the different 
type of Decomposition (\#~13). 
We introduce two inert doublets. 
This allows to have two types of Yukawa interactions for neutrinos:
one is the ordinary one with right-handed neutrinos $\psi_{R}$, 
and the other appears with left-handed component $\psi_{L}$ 
of the SM singlet fermion and violates the 
lepton number.
The particle contents and their charge assignments are summarized 
in Tab.~\ref{Tab:Decom13-charge}. 
The interaction is given by
\begin{align}
\mathcal{L} =& \mathcal{L}_{\rm SM}
\nonumber 
\\
&+
\biggl[
{(Y_{\nu})_{a}}^{\alpha}
\overline{\psi}^{a} {\rm P}_{L}
\eta {\rm i} \tau^{2} L_{\alpha}
+
{(Y_{\nu}')^{a \alpha}}
\overline{\psi^{c}}_{a}
{\rm P}_{L} \eta'^{\dagger} L_{\alpha}
\nonumber 
\\
&\hspace{0.5cm}+
\zeta (H_{1} {\rm i} \tau^{2} H_{2})
(\eta' {\rm i} \tau^{2} \eta)
+
\frac{\lambda}{2}
(\eta^{\dagger} H_{2}) (\eta^{\dagger} H_{2})
\nonumber 
\\
&
\hspace{0.5cm}
+
{\rm H.c.}
\biggr]
\nonumber 
\\
&
+
{M_{a}}^{b} \overline{\psi}^{a} \psi_{b}
+
m_{\eta'}^{2} \eta'^{\dagger} \eta'
+
m_{\eta}^{2} \eta^{\dagger} \eta
-
\mathcal{V}_{\text{scalar}}.
\label{eq:L-Decom13}
\end{align}
The scalar potential is obviously different from 
that of Model A. 
However we assume also 
that it includes the soft violation term of the  
$Z_{5}$ symmetry, which was shown in Eq.~\eqref{eq:Z5soft},
to avoid the problem of the Nambu-Goldstone boson.

\begin{figure}[t]
\unitlength=1cm
\begin{picture}(8,3.5)
\put(0,0){\includegraphics[width=8cm]{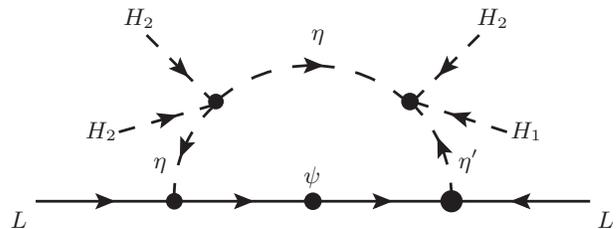}}
\put(0,0){$L$}
\put(7.8,0){$L$}
\put(3.9,0.6){$\psi$}
\put(5.95,0.8){$\eta'$}
\put(1.9,0.8){$\eta$}
\put(4,2.5){$\eta$}
\put(1.5,2.7){$H_{2}$}
\put(1,1.2){$H_{2}$}
\put(6.2,2.7){$H_{2}$}
\put(6.65,1.2){$H_{1}$}
\end{picture}
\caption{Diagram for neutrino masses in 
Model B.
The lepton number is violated at 
the interaction of $\psi$-$L$-$\eta'$,
which is shown with a fat blob.}
\label{Fig:mNu-Decom13}
\end{figure}

With the Lagrangian in Eq.~\eqref{eq:L-Decom13},
neutrino masses are constructed as shown in Fig.~\ref{Fig:mNu-Decom13},
and they are calculated to be
\begin{align}
(m_{\nu})^{\alpha \beta}
=&
-
\lambda \zeta 
\frac{v^{4}}{8}\sin^{2} \beta \sin 2 \beta
\nonumber 
\\
&\times 
\biggl[
(Y_{\nu}'^{{\sf T}})^{\alpha a}
M_{a} \mathcal{I} (x_{a},y)
{(Y_{\nu})_{a}}^{\beta}
\nonumber 
\\
&\hspace{0.5cm}+
{(Y_{\nu}^{\sf T})^{\alpha}}_{a}
M_{a} 
\mathcal{I} (x_{a},y)
(Y_{\nu}')^{a \beta}
\biggr], 
\label{eq:mNu-ModelB}
\end{align}
where $y \equiv m_{\eta'}^{2} /m_{\eta}^{2}$.
The flavour structure of this model is rather involved,
and it cannot be understood with the ordinary seesaw formula
because of two independent 
Yukawa matrices $Y_{\nu}$ and $Y_{\nu}'$.
With the assumption that 
$Y_{\nu}'$ takes the same flavour structure as 
$Y_{\nu}$, Eq.~\eqref{eq:mNu-ModelB} 
is reduced to the ordinary type-I seesaw formula.
Therefore, this model can obviously reproduce 
the mass matrices which are consistent with 
the observed mass squared differences and the mixings. 

The lepton number violating Yukawa interaction 
gives an additional contribution to the LFV process
$\ell_{\alpha} \rightarrow \ell_{\beta} \gamma$.
The decay branching ratio in Model B can be obtained 
by substituting 
\begin{align}
{(\mathcal{C}_{\text{B}})_{e}}^{\mu}
=
{(Y_{\nu}^{\dagger})_{e}}^{a}
\mathcal{F}(x_{a})
{(Y_{\nu})_{a}}^{\mu}
+
(Y_{\nu}'^{\dagger})_{e a}
\mathcal{F}(x_{a})
(Y_{\nu}')^{a \mu},
\label{eq:LFVcoeff-B}
\end{align}
for $\mathcal{C}_{\text{A}}$ in Eq.~\eqref{eq:Brmueg-Decom1}.

\section{Summary and Discussion}
\label{Sec:Summary}

We have proposed the new scenario in which tiny neutrino masses 
are generated via loop-induced $d>$ 5 operators. 
In such a scenario, the scale of tiny neutrino masses can be 
reproduced from the TeV scale physics 
in a natural way without extreme fine tuning because 
the combination of the loop factor $1/(16\pi^2)^{n}$ 
and the additional coefficient of $(v/\Lambda)^{d-5}$ 
provides the sufficient suppression factor. 
We have in particular discussed as examples two concrete models where 
neutrino masses are generated via one-loop induced 
dimension-seven operators due to 
the dynamics of extended Higgs sector and a vector-like Dirac 
neutrino whose mass is assumed to be 
at the TeV scale under the imposed exact discrete $Z_2$ symmetry. 
We have shown that in these models neutrino masses can be reproduced 
and that the neutrino mixing data are also satisfied without contradicting 
the constraint from the LFV data~\cite{Amsler:2008zzb,Adam:2009ci}.

We here give a comment on phenomenological implications 
in these models. However, the detailed discussion is
beyond the scope of this paper, and it is given 
elsewhere~\cite{in-preparation}.
First of all, a common feature of these models is the 
extended Higgs sectors, in which there are two $Z_2$-even 
Higgs doublets and one or two $Z_2$-odd doublets.  
Phenomenology of the THDM has been 
discussed in literature. 
The Higgs potential is constrained by the perturbative 
unitarity~\cite{Lee:1977eg,Kanemura:1993hm,Akeroyd:2000wc,Ginzburg:2003fe}, the vacuum stability~\cite{Deshpande:1977rw,Nie:1998yn,Kanemura:1999xf}, and 
also electroweak precision data~\cite{Lim:1983re,Haber:1992py,Pomarol:1993mu}. 
When the type-II THDM is assumed,
the bounds from $b\rightarrow s \gamma$~\cite{Ciuchini:1997xe}, 
$B \rightarrow \tau \nu$~\cite{Hou:1992sy,Isidori:2006pk,Isidori:2007jw}
and 
the leptonic tau decay~\cite{Krawczyk:2004na} have also to be taken 
care.
The discovery of extra Higgs bosons in addition to the lightest 
(SM-like) Higgs boson and the measurement of their properties 
are important to test these models.
In these models, the induced neutrino masses are multiplied by the
factor of $\sin^{2} \beta \sin 2 \beta$, so that a large value of 
$\tan \beta$ gives a further suppression factor. 
This may bring an interesting correlation between 
neutrino masses and the physics of the Higgs sector. 

The experimental confirmation of the $Z_{2}$ odd sector
is essentially important too.  
%
Especially, the lightest $Z_2$ odd particle can be a 
candidate of dark matter if it is electrically (and colour) neutral. 
In these models, there are two possibilities for the DM candidate; 
i.e., 1) the lightest $\eta^0$ boson is the DM or 
2) the Dirac neutrino $\psi$ is the DM.

In Case 1), the phenomenology of such $Z_2$ odd sector has been 
studied with the physics of the DM candidate in the context of 
the dark doublet model~\cite{Barbieri:2006dq}
and the radiative seesaw 
models~\cite{Krauss:2002px,Ma:2006km,Aoki:2010tf}. 
An interesting signature of DM may be the 
invisible decay of the (SM-like) Higgs boson when DM is 
lighter than a half of the Higgs boson mass. 
It is expected that the branching ratio of the Higgs boson 
invisible decay of greater than 
50 \% (1\%) can be detected at the LHC (at the ILC).  
The direct DM search is also important for the case of 1). 
The multi Higgs portal dark matter has been discussed in 
Ref.~\cite{Aoki:2009pf}.
The detailed comprehensive study for models of the Higgs portal dark matter 
has been done in Ref.~\cite{Kanemura:2010sh} in a specific scenario where
only the Higgs boson and the DM candidate are electroweak scale and the
other new particles are supposed to be decoupled. 
The collider phenomenology of the Higgs sectors with dark doublet fields
has been studied in \cite{Lundstrom:2008ai} at the LEP,
in Ref.~\cite{Cao:2007rm,Dolle:2009fn,Dolle:2009ft,Miao:2010rg}
at the LHC and in Ref.~\cite{Aoki:2010tf} at the ILC. 
For the test of our model, many parts 
of these previous studies can be applied. 

If $\psi$ is dark matter corresponding to 
the case of 2),  the situation may be similar to 
the case in the model by Ma where the right-handed neutrino is 
the DM candidate which
has been studied in detail in Ref.~\cite{Kubo:2006yx}.
However, $\psi$ is a Dirac neutrino, not a Majorana neutrino, 
so that the DM number can be assigned. The DM number may 
be dynamically generated in the context of asymmetric DM. 
Details of these issues are discussed in Ref.~\cite{in-preparation}.

\begin{center}
{\bf Acknowledgments}
\end{center}

This work was supported, in part, by Grant-in-Aid for Scientific Research, 
Japan Society for the Promotion of Science, Nos.~18034004 (C) 
and 22244031 (A).

\bibliography{./loopDim7}
\bibliographystyle{apsrev}

\end{document}